\documentclass[prl,aps,twocolumn,showpacs,floatfix]{revtex4}
\usepackage{mathrsfs}
\usepackage{graphicx}           
\usepackage{dcolumn}            
\usepackage{bm}                 
\usepackage{hyperref}           
\usepackage{hypernat}           
\usepackage[latin1]{inputenc}   
\usepackage{pstricks}           
\usepackage{amsmath,amssymb}
\usepackage{version}
\DeclareGraphicsExtensions{.eps,.ps}

\begin{document}

\title{Attosecond probing of instantaneous AC Stark shifts in helium atoms}

\author{Feng He$^{1,2}$\footnote{fhe@sjtu.edu.cn}, Camilo Ruiz$^3$, Andreas Becker$^4$,
and Uwe Thumm$^5$}

\affiliation{ $^1$Key Laboratory for Laser Plasmas (Ministry of Education) and
Department of Physics, SJTU, Shanghai 200240, People's Republic of China\\
$^2$Max-Planck-Institute f\"ur Kernphysik, Saupfercheckweg 1, D-69117 Heidelberg, Germany\\
$^3$Centro de Laseres Pulsados CLPU, Plaza de la Merced s/n E-37008 Salamanca, Spain\\
$^4$Department of Physics and JILA, University of Colorado,
Boulder 80309-0440, USA\\
$^5$James R. Macdonald Laboratory, Kansas State University,
Manhattan, Kansas 66506, USA }

\date{\today}

\begin{abstract}
Based on numerical solutions of the time-dependent Schr\"odinger equation
for either one or two active electrons, we propose a method for observing
instantaneous level shifts in an oscillating strong infrared (IR) field in time,
using a single tunable attosecond pulse to probe excited states of the
perturbed atom. The ionization probability in the combined fields depends
on both, the frequency of the attosecond pulse and the time delay
between both pulses, since the IR field shifts excited energy
levels into and out of resonance with the attosecond probe pulse. We
show that this method (i) allows the detection of instantaneous
atomic energy gaps with sub-laser-cycle time resolution and (ii) can
be applied as an ultrafast gate for more complex processes such as
non-sequential double-ionization.
\end{abstract}
\maketitle

The energetic shift of atomic levels in external electric
fields is a well-known phenomenon and usually referred to as ``Stark
shift''. For  static fields that are much weaker than intra-atomic
Coulomb fields, Stark shifts can be calculated using perturbation
theory \cite{Bransden03}. For oscillating external fields in the
optical and near-IR range, perturbation theory breaks down at
intensities of $\approx 10^{12}$ W/cm$^2$~\cite{Faisal},
orders of magnitudes below the peak intensities available in
state-of-the-art ultrashort laser laboratories. If such strong
external fields are maintained over many optical cycles,
cycle-averaged level shifts can be evaluated, e.g.\ by exploiting
the quasi periodicity of the external field using the
non-perturbative Floquet
theory~\cite{Chu85,Doerr90,Rottke94}. However, for the
recently developed strong few-cycle IR laser
pulses~\cite{Zhou94,Nisoli97,Durfee99,Apolonski00,Schenkel03},
atomic level shifts are non-perturbative in nature and also
render the continuum-wave Floquet picture inapplicable.

Modern pump-probe experiments combine extended ultraviolet (XUV)
attosecond pulses of sub-IR-cycle pulse lengths (1~as =
10$^{-18}$~s) with phase-coherent IR laser pulses to observe
electronic dynamics in atoms, molecules and solids
\cite{Hentschel01,Itatani02,Drescher02,Uiberacker07,Cavalieri07,Goulielmakis10}.
The role of laser-dressed highly excited energy levels in
atomic excitation and ionization has been studied
recently using attosecond technology \cite{Johnsson07,Ranitovic10}.
In this Letter, we show that the attosecond pump-probe technique
should also enable the measurement of {\em instantaneous} level
shifts of  low lying bound atomic 
states in alternating optical electric fields. Further, we demonstrate how
the control of instantaneous level shifts can be exploited to gate
other  strong-field phenomena, such as non-sequential double
ionization (NSDI).

To this end, we simulate an XUV pump - IR probe scenario. We choose
IR laser fields with negligible distortion of the ground state of
He, that are, however, strong enough to couple low excited and
continuous states, inducing noticeable level splitting, shift, and
decay. For the XUV pulses we fix the number of cycles and vary the
central frequency of the pulse. Key to our investigation is the
observation that, for a given central frequency $\omega_{SA}$  of
the single attosecond (SA) pulse and depending on the delay
$\Delta t$ between pump and probe pulse, the IR pulse may shift
low-lying bound states into or out of resonance with one-photon
excitations from the He ground state. The
excited atom may then be easily ionized by the 
IR pulse. If the SA pulse is applied while the instantaneous
level energies are off (in) resonant with $\omega_{SA}$, less (more)
excitation and thus less (more) ionization out of excited states is
expected to occur. This suggests that detection of the ionization
probability as a function of $\omega_{SA}$ and $\Delta t$ can be used to
track the instantaneous Stark shifts.

We will analyze this strategy by modeling He in the so-called
single-active-electron approximation (SAEA), in which the
time-dependent Schr\"odinger equation (TDSE) in velocity gauge reads
(unless indicated otherwise, we use Hartree atomic units, $e = m =
\hbar = 1$):
\begin{equation}
i\frac{\partial \Psi(z,\rho; t)}{\partial t}=
\left [ \frac{\left(p_z-A(t)/c\right)^2+p_{\rho}^2 }{2}+V(z,\rho) \right ]\Psi(z,\rho;t),
\label{sch}
\end{equation}
where $z$ and $\rho$ are cylindrical electronic coordinates
parallel or perpendicular to the laser-polarization axis,
respectively, and $p_z$ and $p_{\rho}$ are the corresponding
conjugate momentum operators. $A(t)=-c\int dt E(t)$ is the vector
potential associated with the XUV and IR laser
fields and $c$ the speed of light. $V(z,\rho)$ models
electronic correlation in terms of screening of the nuclear Coulomb
potential by the passive electron~\cite{Tong05}.

\begin{figure}
\includegraphics[width=0.85\columnwidth]{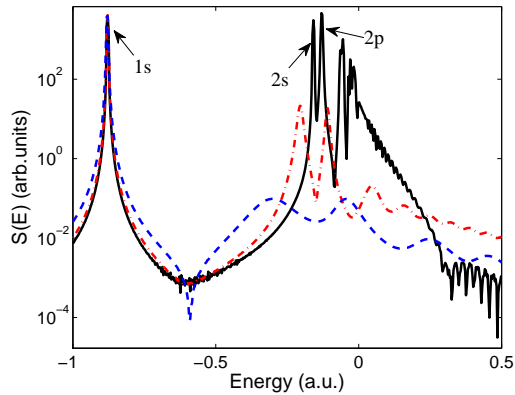}
\caption{(Color online) Energy levels for He, 
calculated in SAEA without (black line) and with static external
electric fields with intensities of $10^{13}$ W/cm$^2$ (red
dash-dotted line) and $10^{14}$W/cm$^2$ (blue dashed line).
 }
\label{level}
\end{figure}

For later reference, we first compute energy-level shifts
in the static field $E_{st}$ in the length gauge. Fig. \ref{level}
shows the spectrum $S(E)$ of He atoms without external field
and in static fields with intensities of $10^{13}$
and $10^{14}$W/cm$^2$,
calculated by Fourier transformation of the autocorrelation
function, $S(E) = |\int dt \langle\psi(t)|\psi(0)\rangle e^{-i E
t}|^2$~\cite{Hermann88}. At these intensities the ground state is
nearly unaffected by the external static field, while the excited
$2s$ and $2p$ states are shifted to lower and higher energies,
respectively~\cite{footnote1}. The level
spacing $E_{1s,2s}^{\text{(static)}}$ between the field-shifted $1s$ and
$2s$ levels decreases with increasing $E_{st}$, while
$E_{1s,2p}^{\text{(static)}}$ increases.

Exposed to the combined electric field of a SA pulse and a
delayed IR pulse $E_{IR}(t)$
\begin{eqnarray}
E(t)
&=&
E_{SA}\sin (\omega_{SA}t)\exp\left
[-2\ln 2 \left (\frac{t-\Delta t}{\tau_{SA}}\right )^2\right ]
\nonumber\\
&&+E_{IR}\sin(\omega_{IR}t)\cos^2\left(\frac{\pi t}{\tau_{IR}}\right
),  \label{laser}
\end{eqnarray}
with $E_{IR}=0$ if $|t| > \tau_{IR}/2$,
level shifts induced by $E_{IR}(t)$ can be probed on an attosecond time scale. We choose Gaussian SA
pulses with an  intensity of $2\times 10^{13}$~W/cm$^2$, a pulse
duration $\tau_{SA}$ (FWHM) of two XUV cycles, and variable
$\omega_{SA}$, and a cos$^2$-shaped IR laser pulse with a central
wavelength of 800 nm, an intensity of $3\times 10^{14}$~W/cm$^2$,
and a pulse duration $\tau_{IR}$ of four IR cycles. We neglect the
spatial intensity profile of the IR laser pulse since SA pulses can
be made to only overlap with the spatial center of the IR pulse
\cite{Takahashi07,Singh10}.

We calculate the ionization probability of He in the oscillating
field, Eq.\ (\ref{laser}), by propagating
Eq.\ (\ref{sch}) on a numerical grid with equidistant
spacings $\Delta z=\Delta\rho=0.3$  and time steps $\delta t=0.05$.
The spatial grid includes 2000 (400) points and covers the range
from $-300$ to 300 (0 to 120) along the $z$ ($\rho$) axis. We use a
$\cos^{1/6}$ masking function to suppress reflections from
the grid boundaries~\cite{He08} and determine the
single-ionization probability from the accumulated outgoing
electronic current into the absorber region. The wave function
is propagated until the ionization probability becomes
stabilized.

\begin{figure}
\includegraphics[width=0.95\columnwidth]{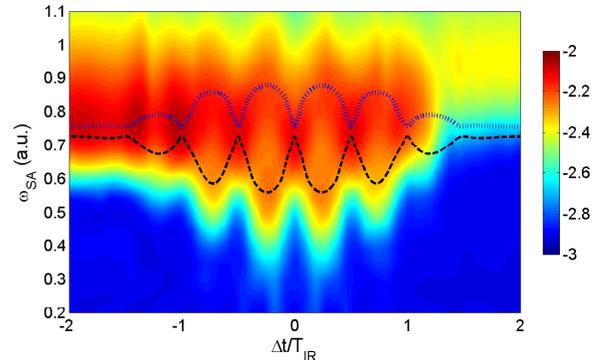}
\caption{(Color online) Ionization probabilities (logarithmic
color/grey scale) of He calculated in SAEA as a function of the
center frequency $\omega_{SA}$ of the SA pulse and  time delay
$\Delta t$ between the SA and IR laser pulses
in units of the IR laser period $T_{IR}$.
Superimposed dashed and dotted curves show the quasi-static level
spacings $E^{\text{(static)}}_{1s,2s}$ and $E^{\text{(static)}}_{1s,2p}$,
respectively. Laser parameters are given in the text.}
\label{fig2}
\end{figure}

\begin{figure}
\includegraphics[width=0.95\columnwidth]{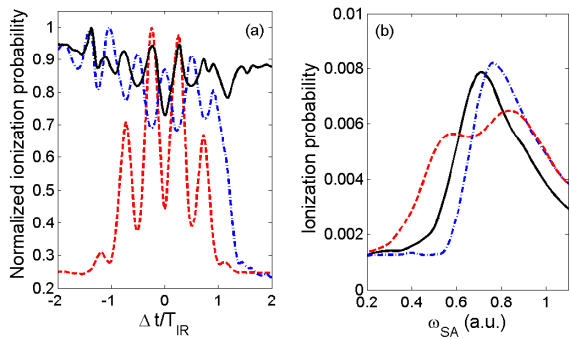}
\caption{(Color online) (a) Ionization probabilities, normalized to
their respective maxima, as a function
of $\Delta t$ for three values of $\omega_{SA}$:
0.5 a.u.~(red dashed curve),
0.75 a.u.~(blue dash-dotted curve), and
1.1 a.u.~(black solid curve). (b) Ionization probabilities as a function of
$\omega_{SA}$ for three values of $\Delta t / T_{IR}$:
$-2$ (blue dash-dotted line),
$-0.25$ (red dashed line), and $0$ (black solid line).
}
\label{fig3}
\end{figure}

The single ionization probability $P_S$
is shown in Fig.~\ref{fig2}. For $\Delta t <0$ the SA pulse
precedes the center of the IR laser pulse. Accounting for the large
bandwidth of the SA pulse, enhanced ionization may occur, even for
$\omega_{SA}$ below the ionization threshold, via excitation to a IR
laser-dressed bound state followed by ionization in the remaining IR
field. Single ionization is thus
mediated by either the high-energy side of the XUV-pulse spectrum
that extends above the ionization continuum
or in a two-step process via transiently excited atoms. Since
$P_S(\omega_{SA},\Delta t)$ behaves differently for different
$\omega_{SA}$, we analyze separately three $\omega_{SA}$ intervals.
In order to support our interpretations we present in
Fig.~\ref{fig3} (normalized) ionization probabilities as functions
of (a) $\Delta t$ and (b) $\omega_{SA}$ fixing the other parameter
to different values.

(1) For $\omega_{SA}<0.6$ and  negligible pulse-overlap ($|\Delta
t|>1.5 T_{IR}$, where $T_{IR}$ is the period of the IR field), there
is no enhancement of the ionization probability. However, $P_S$ is
enhanced and oscillates in the pulse-overlap  region ($|\Delta
t|<1.5T_{IR}$) as a function of $\Delta t$ with period $T_{IR}/2$.
$P_S$ reaches maxima when the SA pulse is applied at maximal IR
electric-field magnitudes $|E_{IR}|$ (Fig.~\ref{fig3}(a), red-dashed
line). We explain this behavior as follows. For $|\Delta
t|>1.5T_{IR}$, the energy gap between the (essentially) field-free
$1s$ and $2p$ level, $E_{1s,2p}^{\text{(free)}}=0.75$, is
larger than $\omega_{SA}$, and the spectral width of the SA
pulse does not overlap with the $2p$ level.
Thus excitation by the SA pulse is negligible. In contrast,
for $|\Delta t|<1.5T_{IR}$ the  dressing by the IR-laser field
repeatedly shifts $E_{1s,2s}^{\text{(dressed)}}$ into resonance with
larger values of $\omega_{SA}$,
allowing for efficient excitation followed by ionization in the IR
field. The $T_{IR}/2$ oscillation of  $P_S$ indicates that
$E_{1s,2s}^{\text{(dressed)}}$ depends on the {\em instantaneous} IR
intensity.

(2) If $\omega_{SA}$ approximately matches the field-free energy gap
$E_{1s,2p}^{\text{(free)}}$
and $\Delta t < -1.5T_{IR}$, excitation to the $2p$ state by the SA
pulse followed by subsequent ionization via the IR pulse is very
likely. In contrast, the IR pulse by itself is not able to excite
and leaves the atom in its ground state. Thus, for $\Delta
t>1.5T_{IR}$, the subsequent not-IR-assisted SA pulse can only
excite, but not ionize, the atom. In the overlap region ($|\Delta
t|<1.5T_{IR})$ a small-amplitude IR-laser-induced oscillation
of $P_S$ occurs. Maxima in $P_S$ occur at vanishing instantaneous
IR intensity (Fig.~\ref{fig3}(a), blue dashed-dotted line),
since the energy gap $E_{1s,2p}^{\text{(dressed)}}$ increases with
field strength which leads to a mismatch with $\omega_{SA}$. In
Fig.~\ref{fig3} the blue dashed-dotted line and red dashed line have
opposite carrier phases, since the corresponding energy gaps
are shifted into and out of resonance with the SA frequency,
asynchronously.

(3) For $\omega_{SA}>0.8$ the spectrally broad SA pulse can directly ionize He from the ground
state, leading to noticeable ionization probabilities even without pulse
overlap (Fig.~\ref{fig3}(a), black solid line). The larger
ionization probability for $\Delta t<-1.5 T_{IR}$ is due to dominant
SA-pulse excitation followed by IR ionization, while this two-step
mechanism does not apply for $\Delta t>1.5 T_{IR}$. In the overlap
region $|\Delta t|<1.5 T_{IR}$, the red dashed and black solid lines
are in phase due to synchronous shifts into and out of resonance
with $E_{1s,2s}^{\text{(dressed)}}$ and
$E_{1s,2p}^{\text{(dressed)}}$, respectively.

Further evidence for the observation of $P_S(\omega_{SA},\Delta t)$
allowing the detection of instantaneous level shifts is obtained by
comparison with the instantaneous quasi-static energy gaps
$E_{1s,2s}^{\text{(static)}}(t)$ and
$E_{1s,2p}^{\text{(static)}}(t)$ in the static external field
$E_{st}=E_{IR}(t)$, where $E_{IR}(t)$ is the instantaneous electric
field of the IR-laser pulse at time $t$ (cf. Fig.~\ref{level}). The
results are drawn as dashed and dotted curves in Fig.~\ref{fig2} and
match the contours of the ionization probability.

For a quantitative investigation, we present the ionization probability
$P_S$ in Fig.~\ref{fig3}(b) as a function of $\omega_{SA}$ at
three different delays. At $\Delta t = -T_{IR}/4 $ (corresponding to
a maximum of $|E_{IR}(t)|$, red dashed line) the
instantaneous energy spacings can be identified by the {\em two}
local extrema in $P_S$.  The positions of  these maxima as a
function of $\omega_{SA}$ match
the corresponding quasi-static energy gaps
$E_{1s,2s}^{\text{(static)}}(t)$ and
$E_{1s,2p}^{\text{(static)}}(t)$ for $E_{st} = |E_{IR}(t)|$ at
$t=-T_{IR}/4$.

In contrast, application of the SA
pulse at a zero of the IR field ($\Delta t =0$, black solid line)
leads to just one maximum in the ionization probability. The
position of this maximum as a function of $\omega_{SA}$ is slightly
shifted to lower frequencies as compared to the field-free energy
spacing $E_{1s,2p}^{\text{(free)}}$ (blue dash-dotted line for
$\Delta t = -2T_{IR}$ in Fig.~\ref{fig3}(b)). This is due to the
fact that the SA pulse has a finite non-zero temporal width and
probes the (field-dressed) $2s$ and $2p$ levels near a zero
of the IR field as well.

Understanding of the ultrafast shift of the electronic levels can be
useful to control electronic transitions with  a high temporal
resolution. Single ionization of an atom (or a molecule) is the
initial step in many strong-field processes, such as higher-order
harmonic and attosecond-pulse generation, NSDI,
or molecular dissociation. Our findings above suggest that these processes can be gated
in a controlled way on an attosecond time scale. We demonstrate the
potential of this gating technique for NSDI
of He by solving the TDSE
\begin{equation}
\label{eq:2e}
 i\frac{\partial}{\partial t}\Psi(Z,\overline{z},\overline{\rho}; t) = \left
[H_0 -\frac{P_Z\cdot A(t)}{c}\right ]
\Psi(Z,\overline{z},\overline{\rho}; t),
\end{equation}
within a correlated two-electron model~\cite{Ruiz06}, with the
field-free Hamitonian
\begin{equation}
H_0= \frac{P_Z^2}{4} + p_{\overline{\rho}}^2 + p_{\overline{z}}^2
+\frac{1}{\sqrt{\overline{\rho}^2+\overline{z}^2}}-
\sum_{j=1,2}{\frac{2}{\sqrt{r_{j}^2+s}}}.
\end{equation}
where $r_j^2 = (Z+(-1)^j \overline{z}/2)^2+\overline{\rho}^2/4$. In
this model the electronic center-of-mass coordinate $Z$ is
constrained along the polarization direction, $\overline{z}$
and $\overline{\rho}$ represent the coordinates of the
(unconstrained) relative motion of the electrons parallel and
perpendicular to the polarization axis, respectively. $P_Z$,
$p_{\overline{z}}$ and $p_{\overline{\rho}}$ are the corresponding
conjugate momentum operators. The soft core
parameter $s = 0.135$. We solve Eq.~(\ref{eq:2e}) on a
three-dimensional numerical grid with equidistant spacing $\Delta
Z=\Delta \overline{z}=\Delta\overline{\rho}=0.3$ and 600, 1200, and
200 grid points along the $Z$, $\overline{z}$, and $\overline{\rho}$
axis, respectively. To quantify the probabilities of single and
double ionization, we partition the grid as $r_1 <12$ and $r_2< 12$
(neutral helium), $r_i < 6$ and $r_j > 12$ ($i \ne j = 1,2$,
He$^+$) and the complementary space (He$^{2+}$) (c.f.
\cite{Ruiz06}).

\begin{figure}
\includegraphics[width=0.95\columnwidth]{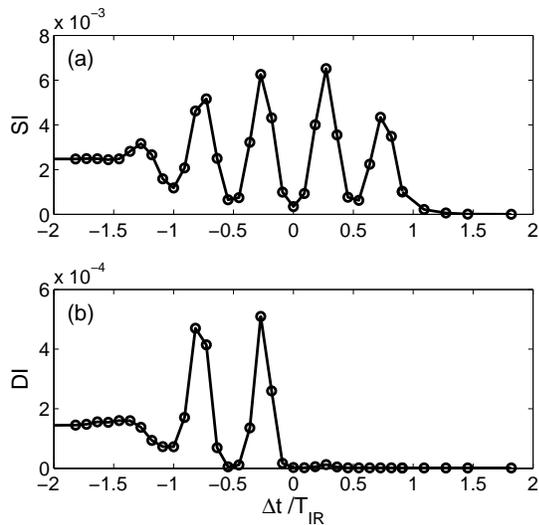}
\caption{Probabilities for single ionization (SI)(a) and double ionization (DI) (b) as a
function of the time delay between SA and IR pulses. The SA pulse
has a central frequency of $\omega_{SA} =0.76$ a.u.
and a peak intensity of $2\times
10^{13}$W/cm$^2$. The IR pulse has a central wavelength of 800 nm and a
peak intensity of $3\times 10^{14}$W/cm$^2$. The simulation results
(circles) are interpolated by lines.} \label{double}
\end{figure}

Fig. \ref{double}(a) and (b) show the single and double ionization
probabilities as functions of $\Delta t$ between the SA and IR-laser
electric fields  (\ref{laser}). The SA (IR) pulse has
a central wavelength of 60 (800) nm and peak intensity $2\times
10^{13}$ ($3\times 10^{14}$) W/cm$^2$. The value of $\omega_{SA} =
0.76$ is chosen such that it is repeatedly in resonance with the
field-dressed $1s2s$ level. Note that
$E_{1s,2s}^{\text{(free)}}=1$ in this two-electron model.
Therefore, Fig.\ref{double} (a) closely resembles in shape the
$\omega_{SA}<0.6$ result in the SAEA calculation in Fig.
\ref{fig3}(a), and
the single-ionization-probability maxima occur when the SA pulse
coincides with the maxima of $|E_{IR}(t)|$.

Once released, an electron can be accelerated and driven back to the
core by the IR laser field, causing non-sequential rescattering
ionization of the ionic core after about $2/3\ T_{IR}$
\cite{Corkum93}. Hence, the single and double-ionization
probabilities oscillate with the same period, $T_{IR}/2$
(Fig.~\ref{double}), with maxima at almost identical delays. The
double ionization probability almost vanishes
for $\Delta t>0$, since the decreasing IR laser intensity does not
transfer sufficient energy to the rescattering electron to ionize
the He$^+$ core.
If the SA appears much earlier than the IR pulse ($\Delta t <
-1.5T_{IR}$),
the SA pulse first excites
the helium atom, which is later singly and doubly ionized, once
nearly maximal IR intensities
are reached \cite{Chen10}.

In conclusion, our simulations indicate that by applying a SA pulse
{\it instantaneous} energetic shifts in singly excited He induced by
a strong few-cycle IR laser pulse are mapped onto oscillations in
delay-depended single ionization  probabilities. We interpret these
oscillations in terms of a two-step process, where excitation in the
SA pulse is followed by efficient ionization out of excited states
in the IR field. The excited-state population depends on the
instantaneous energy gaps between the ground and excited
states in the IR-laser field. Knowledge of the instantaneous energy levels
in the strong IR field may lead to new schemes for the coherent
control of NSDI, high harmonic generation, and molecular
dissociation.

This work was supported by the NSF of China (Grant No. 10734130)
and National Basic Research Program of China (Grant No. 2007CB310406, 2007CB815105), 
the US NSF, the Division of Chemical Sciences, Office of
Basic Energy Sciences, Office of Energy Research, US~DOE,
and by the MEC FIS 2009- and the Ram\'on y Cajal Research
grants.


\begin{thebibliography}{99}
%
\bibitem{Bransden03}
B.H. Bransden and C.J. Joachain, Physics of atoms and
molecules (Pearson Education, Ltd, Singapore, 2003).

\bibitem{Faisal}
F. H. M. Faisal,
{\it Theory of Multiphoton Processes}
(Plenum Press, New York, 1987).
%
\bibitem{Chu85}
Shih-I. Chu and J. Cooper,
\pra \textbf{32}, 2769 (1985).
%
\bibitem{Doerr90}
M. D\"orr, R.M. Potvliege, and R. Shakeshaft,
\pra\textbf{41}, 558 (1990).
%
\bibitem{Rottke94}
H. Rottke {\it et al.},
\pra \textbf{49}, 4837 (1994).
%
\bibitem{Zhou94}
J. Zhou {\it et al.},
Opt. Lett. \textbf{19}, 1149 (1994).
%
\bibitem{Nisoli97}
M. Nisoli {\it et al.},
Opt. Lett. \textbf{22}, 522 (1997).
%
\bibitem{Durfee99}
C.G. Durfee, S. Backus, H.C. Kapteyn, and M.M. Murnane,
Opt. Lett. \textbf{24}, 697 (1999).
%
\bibitem{Apolonski00}
A. Apolonski {\it et al.},
Phys. Rev. Lett. \textbf{85}, 740 (2000).
%
\bibitem{Schenkel03}
B. Schenkel {\it et al.},
Opt. Lett. \textbf{28}, 1987 (2003).
%
\bibitem{Hentschel01}
M. Hentschel {\it et al.},
Nature (London) \textbf{414}, 509 (2001).
%
\bibitem{Itatani02}
J. Itatani {\it et al.},
\prl \textbf{88}, 173903 (2002).
%
\bibitem{Drescher02}
M. Drescher {\it et al.},
Nature (London) \textbf{419}, 803 (2002).
%
\bibitem{Uiberacker07}
M. Uiberacker {\it et al.},
Nature (London) \textbf{446}, 627 (2007).
%
\bibitem{Cavalieri07}
A. Cavalieri {\it et al.},
Nature (London) \textbf{449}, 1029 (2007).
%
\bibitem{Goulielmakis10}
E. Goulielmakis {\it et al.},
Nature (London) \textbf{466}, 739 (2010).
%
\bibitem{Johnsson07}
P. Johnsson {\it et al.},
\prl \textbf{99}, 233001 (2007).
%
\bibitem{Ranitovic10}
P. Ranitovic {\it et al.},
New J. Phys. \textbf{12}, 013008 (2010).
%
\bibitem{Tong05}
X. M. Tong and C. D. Lin,
J. Phys. B: At. Mol. Opt. Phys. \textbf{38} 2593(2005).
%
\bibitem{Hermann88}
M.R. Hermann and J.A. Fleck Jr, \pra \textbf{38}, 6000 (1988).
%
\bibitem{footnote1}
We designate dressed states by $nl$, following~\cite{Doerr90},
to indicate their undressed origins.
%
\bibitem{Takahashi07}
E.J. Takahashi {\it et al.}, \prl \textbf{99}, 053904 (2007).
%
\bibitem{Singh10}
K.P. Singh {\it et al.}, \prl \textbf{104}, 023001 (2010).
%
\bibitem{He08}
F.\ He, A.\ Becker, and U.\ Thumm, \prl \textbf{101}, 213002 (2008).
%
\bibitem{Ruiz06} C. Ruiz, L. Plaja, L. Roso, and A. Becker,
\prl \textbf{96}, 053001 (2006).
%
\bibitem{Corkum93}
P.B. Corkum, \prl \textbf{71}, 1994 (1993).
%
\bibitem{Chen10}
S. Chen, C. Ruiz, and A. Becker,
\pra {\bf 82}, 033426 (2010).
%
\end{thebibliography}
\end{document}